\documentclass[%
 reprint,
 superscriptaddress,
 nofootinbib,
 amsmath,amssymb,
 aps,
 prd,
 floatfix,
]{revtex4-2}
\usepackage{float}
\usepackage{graphicx}
\usepackage{dcolumn}
\usepackage{bm}
\usepackage[dvipsnames]{xcolor}
\usepackage{multirow}

\usepackage[caption=false]{subfig}

\usepackage{hyperref}
\hypersetup{colorlinks = true,
            linkcolor = blue,
            urlcolor  = blue,
            citecolor = blue,
            anchorcolor = blue}

\usepackage{soul}
\usepackage{comment}
\usepackage{siunitx}
\newcommand{\SIadj}[2]{\SI[number-unit-product={\text{-}}]{#1}{#2}}
\usepackage{orcidlink}

\newcommand{\Yale}{\affiliation{Department of Physics, Yale University, New Haven, Connecticut 06520, USA}}
\newcommand{\WrightLab}{\affiliation{Wright Laboratory, Yale University, New Haven, Connecticut 06520, USA}}
\newcommand{\YQI}{\affiliation{Yale Quantum Institute, Yale University, New Haven, Connecticut 06520, USA}}
\newcommand{\Iceland}{\affiliation{Science Institute, University of Iceland, 107 Reykjavik, Iceland}}
\newcommand{\NewUU}{\affiliation{New Uzbekistan University, Tashkent, Uzbekistan}}
\newcommand{\Stockholm}{\affiliation{The Oskar Klein Centre, Department of Physics, Stockholm University,
AlbaNova, SE-10691 Stockholm, Sweden}}
\newcommand{\ORNL}{\affiliation{Physics Division, Oak Ridge National Laboratory, Tennessee 37831, USA}}
\newcommand{\Berkeley}{\affiliation{Department of Nuclear Engineering, University of California, Berkeley, California 94720, USA}}
\newcommand{\BerkeleyPhysics}{\affiliation{Department of Physics, University of California, Berkeley, California 94720, USA}}
\newcommand{\ASU}{\affiliation{Department of Physics, Arizona State University, Tempe, Arizona 85287, USA}}
\newcommand{\Hopkins}{\affiliation{Department of Physics and Astronomy, The Johns Hopkins University, Baltimore, Maryland 21218, USA}}
\newcommand{\Wellesley}{\affiliation{Wellesley College, Wellesley, Massachusetts 02481, USA}}
\newcommand{\Cambridge}{\affiliation{Institute of Astronomy and Kavli Institute for Cosmology, University of Cambridge, Madingley Road, Cambridge, CB3 0HA, UK}}
\newcommand{\Cavendish}{\affiliation{Cavendish Laboratory, Department of Physics, University of Cambridge, JJ Thomson Avenue, Cambridge, CB3 0HE, UK}}

\newcommand{\Colorado}{\affiliation{Department of Physics, University of Colorado, Boulder, Colorado 80309, USA}}
\newcommand{\MIT}{\affiliation{Center for Theoretical Physics, MIT, Cambridge, Massachusetts 02139, USA}}
\newcommand{\Fermilab}{\affiliation{Fermi National Accelerator Laboratory, Batavia, Illinois 60510, USA}}
\newcommand{\TDLee}{\affiliation{T.D.~Lee Institute and Wilczek Quantum Center, Shanghai Jiao Tong University, Shanghai 200240, China}}

\begin{document}


\title{Design of ALPHA Phase I: A Plasma Haloscope for 10--20\,GHz Post-Inflation Axions}

\author{Xiran~Bai}
\Yale\WrightLab

\author{Rustam~Balafendiev}
\Iceland

\author{Sean~E.~Barrett}
\Yale\YQI

\author{Eunice~Beato}
\Yale\WrightLab

\author{Pavel~Belov}
\NewUU

\author{Charles~D.~Brown}
\Yale\WrightLab\YQI

\author{Eduardo~A.~Castro~Muñoz}
\Yale\WrightLab

\author{Jan~Conrad}
\Stockholm

\author{Marcel~Demarteau}
\ORNL

\author{Alex~Droster}
\Berkeley

\author{Joseph~Dubois}
\ASU

\author{Jonathan~Echevers}
\Berkeley

\author{Ali~Elhadi}
\Yale

\author{Jim~Enriquez}
\NewUU

\author{Maryam~Haytham~Esmat}
\Hopkins

\author{Andrea~Gallo~Rosso}
\Stockholm

\author{Eleanor~Graham}
\Yale\WrightLab

\author{Chloe~Greenstein}
\Hopkins

\author{Jon~E.~Gudmundsson}
\Iceland\Stockholm

\author{Karsten~M.~Heeger}
\Yale\WrightLab

\author{Ishaan~Iyer}
\Berkeley\BerkeleyPhysics

\author{Heather~Jackson}
\Berkeley

\author{Junu~Jeong}
\email[\textbf{Corresponding author: }]{jun-woo.jeoung@fysik.su.se}
\Stockholm

\author{Michael~J.~Jewell}
\altaffiliation{Currently at IBM}\Yale\WrightLab

\author{Tyler~Johnson}
\Yale\WrightLab

\author{Shriram~Jois}
\Berkeley

\author{Gagandeep~Kaur}
\Stockholm

\author{Claire~Laffan}
\Yale\WrightLab


\author{K.W.~Lehnert}
\Yale\YQI

\author{Samantha~M.~Lewis}
\Wellesley

\author{Jacob~Lindahl}
\Stockholm

\author{Reina~H.~Maruyama}
\Yale\WrightLab\YQI

\author{Philip~Mauskopf}
\ASU

\author{Andrew~M.~Meyer}
\Stockholm

\author{Alexander~J.~Millar}
\altaffiliation{Currently unaffiliated}\Fermilab

\author{Dylan~R.~Miller}
\Berkeley

\author{Hiranya~V.~Peiris}
\Cambridge\Cavendish

\author{Jianyang~Qi}
\Stockholm

\author{Elizabeth~P.~Ruddy}
\Colorado

\author{Sharada~Sahoo}
\Berkeley

\author{Denis~Sakhno}
\NewUU

\author{Michael~Sekatchev}
\Berkeley

\author{Max~Silva-Feaver}
\email[\textbf{Corresponding author: }]{maximiliano.silva-feaver@yale.edu}
\Yale\WrightLab

\author{Shenyang~Shi}
\Yale\WrightLab

\author{Aarav~M.~Sindhwad}
\altaffiliation{Currently at Yale University}\Berkeley

\author{Gaganpreet~Singh}
\Stockholm

\author{Sukhman~Singh}
\Yale\WrightLab

\author{Danielle~H.~Speller}
\Hopkins

\author{Dajie~Sun}
\altaffiliation{Currently at the University of Idaho}\Berkeley

\author{Noshin~Tabassum}
\Yale

\author{Karl~van~Bibber}
\Berkeley

\author{Yongqi~Wang}
\Hopkins

\author{Frank~Wilczek}
\ASU\TDLee\Stockholm\MIT

\author{Mackenzie~Wooten}
\Berkeley

\author{Sabrina~Zacarias}\
\altaffiliation{Currently at UC Santa Barbara}\Yale\WrightLab



\collaboration{ALPHA Collaboration}
\noaffiliation

\date{\today}

\begin{abstract}
The axion is a well-motivated hypothetical particle capable of resolving both the strong CP problem and the dark matter mystery, with recent post-inflationary cosmological simulations favoring masses above \SI{40}{\micro\eV}.
Plasma haloscopes serve as a promising experimental approach to reach theoretically preferred sensitivities in this mass range.
ALPHA, hosted at Yale Wright Laboratory, is an international collaboration developing plasma haloscopes to search for QCD dark matter axions.
In this letter we present the detailed design and sensitivity projection for the first phase of the ALPHA experiment, which will search the mass range from \SIrange{10}{20}{\GHz}~($\sim$\SIrange{40}{80}{\micro \eV}).
This search will make use of wire-array plasma resonators to decouple the physical size from the resonant frequency, a limitation typically faced by traditional microwave cavities, allowing broadband sensitivity approaching KSVZ coupling strengths. 


\end{abstract}

\maketitle

\section{Introduction}

Cosmological and astrophysical observations over almost a century have accumulated evidence for a component of the Universe, dubbed dark matter, that interacts only feebly with ordinary matter if at all~\cite{Bertone:RMP:2018}.
The axion, originally proposed to solve the strong CP problem~\cite{PecceiQuinn:PRL:1977,PecceiQuinn:PRD:1977,Weinberg:PRL:1978,Wilczek:PRL:1978} in quantum chromodynamics (QCD), also offers a natural solution to the dark matter conundrum~\cite{Preskill:PLB:1983,Abbott:PLB:1983,Dine:PLB:1983}.
However, after nearly half a century of investigation, the axion has proven to be elusive.
While theory does not predict a specific mass or coupling coefficient for the axion, QCD axions that solve the strong CP problem are constrained to a range of couplings at each mass~\cite{DiLuzio:2020wdo}, with the KSVZ~\cite{kim1979KSVZ,shifman1980KSVZ2} and DFSZ axion models~\cite{dine1981DFSZ,zhit1980DFSZ2} serving as useful benchmarks.
In addition, if Peccei-Quinn symmetry breaking occurs after inflation, the mass of the axion can in principle be calculated from the present-day dark matter density.
These calculations favor axion masses exceeding \SI{40}{\micro eV} ($\sim\SI{10}{\GHz}$)~\cite{Saikawa:2024bta,Kim:2024wku,Buschmann2022,Benabou:2024msj,Gorghetto:2020qws}, with recent results predicting a mass from \SIrange{45}{450}{\micro eV}~\cite{Buschmann2022,Benabou:2024msj,Saikawa:2024bta}.

To probe this theoretically favored parameter space, resonant searches in strong magnetic fields~\cite{Sikivie:1983:PRL} represent one of the most promising and accessible methods for detecting QCD dark matter axions. A haloscope applies a strong magnetic field along the axis of a resonant structure; the oscillating axion field then sources an oscillating electric field. When the resonant frequency matches that of the dark matter axion, the conversion power is amplified by the quality factor of the resonator. The axion conversion appears as excess power at the resonator's output antenna and is detected with low-noise radio techniques.

Because the exact mass of the dark matter axion is unknown, the resonator frequency must be continuously tuned to match it.
Therefore, scanning is essential, and the corresponding scanning rate has the following proportionality~\cite{Sikivie:1985:PRD,haystac_pI_design,Simanovskaia2023}:
\begin{widetext}
\begin{equation}
\label{eq:scanrate}
\frac{d\nu}{dt} = 2.7\,\mathrm{GHz/year} \left(\frac{\left| C_{a\gamma} \right| }{1.92} \right)^{4} \left(\frac{\rho_{a}}{0.45\,\mathrm{GeV/cm^{3}}} \right)^{2} \left(\frac{\langle \mathbf{B}_{\mathrm{ext}}^{2} \rangle}{\left(9\,\mathrm{T}\right)^{2}} \right)^{2} \left(\frac{V}{0.012\,\mathrm{m^{3}}} \right)^{2} \left(\frac{C}{0.4} \right)^{2} \left(\frac{Q}{1.3\times 10^{4}} \right) \left(\frac{1.32}{N_{\mathrm{sys}}} \right)^{2}.
\end{equation}
\end{widetext}
Here, $C_{a\gamma}$ is the axion-photon coupling coefficient ($C_{a\gamma} \approx -1.92$ for the KSVZ model), $\rho_{a}$ is the local dark matter density, and $\langle \mathbf{B}_{\mathrm{ext}}^{2} \rangle$ is the mean squared external magnetic field within the resonator volume $V$.
$Q$ is the unloaded quality factor of the resonator, $N_{\mathrm{sys}}$ is the total number of system noise photons, and $C$ is the form factor describing how well the electric field of the resonant mode is aligned with the external magnetic field, defined by the following formula:
\begin{equation}
    C = \frac{\left|\int_{V} \mathbf{E} \cdot \mathbf{B}_{\mathrm{ext}} dV \right|^{2}}{\int_{V} \left|\mathbf{B}_{\mathrm{ext}}\right|^{2} dV \int_{V} \varepsilon \left| \mathbf{E}\right|^{2} dV},
\end{equation}
where $\mathbf{E}$ is the mode electric field, $\mathbf{B}_{\mathrm{ext}}$ is the external magnetic field, and $\varepsilon$ is the dielectric constant of the materials inside the resonator.
The product $C \times V$ is considered the effective interaction volume for axion-to-photon conversion.

Maintaining a large effective interaction volume, however, becomes severely challenging at these high predicted masses.
Because the condition for resonant conversion of an axion to a single photon is $h\nu = m c^2(1+ \beta^2/2)$---where the boost factor $\beta \sim O(10^{-3})$ for virialized axions in the Galactic halo---higher axion masses directly  dictate scanning at higher resonant frequencies.
In conventional cavity haloscopes, this requirement leads to a steep loss of sensitivity, as the experiment's scanning rate scales approximately as $\nu^{-6}$ due to the shrinking resonator volume~\cite{Simanovskaia2023,Jeong:NIMS:2023}.

While multiple cavities within the magnet volume can in principle be phase-combined and tuned in concert~\cite{Hangmann:RSI:1990}, engineering challenges will ultimately set practical limits to this tactic.
Consequently, many alternative approaches have been proposed, including multiple-cell cavities~\cite{Jeong:PLB:2018,RADES:JCAP:2018,Chao-Lin:PRD:2025}, higher-order cavities~\cite{ADMX_SideCar:PRL:2018,Kim:JPG:2020,QUAX:PRD:2022,CAPP:PRD:2023}, and dielectric haloscopes~\cite{MADMAX:PRL:2017,Baryakhtar:PRD:2018,DALI:JI:2024}.
Lawson {\it et al.}\ have proposed a different solution to the volume-loss problem, namely the use of a wire-array metamaterial whose plasma frequency can be designed and tuned~\cite{Lawson:2019brd}; unlike a microwave cavity in which the frequency is determined by its overall boundaries and thus physical size, the plasma frequency of a metamaterial is a bulk property based on the geometry of its unit cell.
This decouples the resonant frequency from the macroscopic dimensions, enabling the design of resonators operating at much higher frequencies without sacrificing volume or the scanning rate.

The theory of wire media metamaterials has been elaborated, see e.g.~\cite{Pendry_1998,Belov:PhysRevB:2003}, and simulations performed of wire-array filled resonators for the axion haloscope application~\cite{Balafendiev:PRB:2022}. 
More recently, extensive measurements of wire array metamaterials have been carried out~\cite{Wooten:AdP:2024,Kowitt:PRAppl:2023}, validating the semianalytic theory of Ref.~\cite{Belov:PhysRevB:2003} as an accurate tool to guide resonator design. Based on all these developments, it was possible to articulate a conceptual design for an actual plasma haloscope~\cite{ALPHA:2022rxj}. In this paper, we present the detailed technical design of the ALPHA (Axion Longitudinal Plasma HAloscope) experiment and its projected sensitivity.
ALPHA is a dedicated axion haloscope currently under construction, designed to employ tunable wire‑array metamaterial resonators to search for high‑mass axion dark matter.

The ALPHA program follows a staged experimental approach. Phase~I, which is the primary focus of this work, targets the \SIrange{10}{20}{\GHz} range using plasma resonators together with established near‑quantum‑limited microwave amplifiers.

In Section~\ref{sec:exp}, we describe the overall experimental setup for ALPHA Phase I, which includes the cryostat and magnet, the tunable resonator, the quantum readout system, and the data acquisition chain.
Section~\ref{sec:projection} details the analysis plan alongside the projected sensitivities of the ALPHA experiment.
Finally, Section~\ref{sec:discussion} discusses alternative signal-readout strategies for even higher frequencies, as well as research and development on superconducting cavities.

\section{Experimental Design}
\label{sec:exp}
\begin{figure*}[t!]
    \centering
    \includegraphics[width = \linewidth]{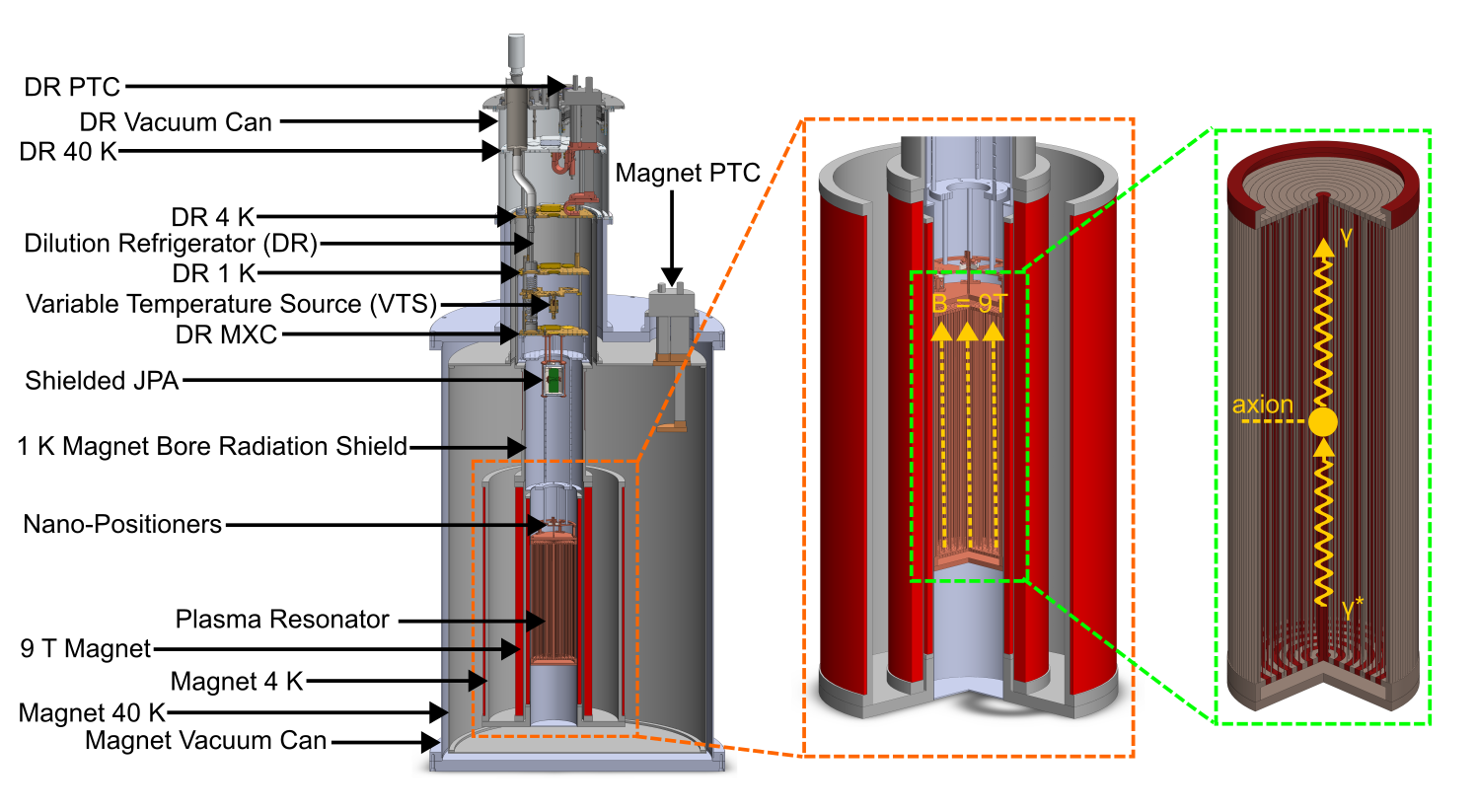}
    \caption{The experimental apparatus. (left) A coupling flange joins a dilution refrigerator (DR) with a superconducting magnet cryostat. The plasma resonator, which confines the axion detection region, is suspended in a magnetic field with magnitude $B_{\mathrm{cent}}= 9~\mathrm{T}$, via a gantry affixed to the mixing chamber (MXC) of the DR. The magnet is cooled to $<$4~K by a pulse tube cryocooler (PTC) and is suspended inside of a $<$40~K radiation shield cooled by a second cooling stage of the PTC. To shield the plasma resonator from 4~K photons from the bore of the magnet a $<$1~K radiation shield cooled by the still of the DR extends into the bore of the magnet. Nano-positioners provide fine position control of the resonator tuning structure within the magnet bore. The magnet windings produce a low-field region above the main magnet bore within which a magnetically shielded Josephson parametric amplifier (JPA) is located as part of the readout chain. A variable temperature source (VTS) allows for noise calibrations. (center) A cutaway of the magnet bore shows 3 concentric magnet windings in red used to 1) produce a uniform 9~T field within the inner 50~cm of the 100~cm bore, 2) reduce the radial extent of the field to minimize equipment and human safety concerns outside the magnet cryostat, and 3) to create the low-field region for the JPA. (right) A cutaway of the plasma resonator in the center of the magnetic cryostat shows schematically the axion conversion scenario where the bath of local dark matter halo axions interacts with the virtual photons, $\gamma^*$, created by the large magnetic field to produce a real photon, $\gamma$, oscillating at the frequency corresponding to the axion mass. If the photon frequency matches the resonator's frequency then an excess of power is detected with low noise radio techniques.}
    \label{fig:experiment_design}
\end{figure*}

The ALPHA Phase~I experiment will be deployed at Yale University’s Wright Laboratory.
At the core of the experiment is a multi-wire tunable resonator immersed in a magnetic field of a wide-bore \SI{9}{\tesla} magnet, operated at cryogenic temperatures and read-out by a set of flux-pumped Josephson parametric amplifiers (JPAs)~\cite{lehnert_jpa}.
A diagram with key components of the experiment is shown in Figure~\ref{fig:experiment_design}. 
With the primary experimental infrastructure complete and pending the arrival of the magnet, the collaboration has integrated all existing components.
The system is currently being validated by conducting a dark photon search with a prototype resonator.

\subsection{Cryostat and Magnet}

The conversion of axions into a microwave signal will be facilitated by a custom cryogen-free magnet, manufactured by Cryogenic Limited, which consists of a superconducting solenoid operating in persistent mode.  The magnet will achieve a maximum central field magnitude $B_{\mathrm{cent}}=\SI{9}{T}$ in a cylindrical bore volume with a diameter of \SI{202}{mm} and a length of \SI{1000}{mm}. Throughout the central \SI{202}{mm} $\times$ \SI{500}{mm} of the bore, the field will be homogeneous to within $\approx$~\SI{3}{\percent}, with a decay rate of $<$\SI{0.1}{ppm\per hr}~($<$\SI{0.01}{ppm\per hr}) for short~(long) persistent-mode timescales. The remaining 500 mm of the bore, 250 mm above and below the homogeneous region, falls outside this specification. Future extensions of ALPHA may make use of this additional space; however, simulations and field-uniformity measurements over the extended region would need to be performed at that time. The magnet will be cooled by pulse tube cryocoolers to maintain an operating temperature $<$\SI{4}{\kelvin}.

The resonator will be suspended within the central region of the magnet’s bore via a support gantry made of oxygen-free high conductivity (OFHC) copper plated stainless steel (SS).
SS is used to reduce damage from potential magnet quenches that induce eddy currents in the gantry.
This gantry will be thermally anchored to the mixing chamber of a Bluefors LD400 dilution refrigerator capable of supplying $\geq$\SI{400}{\micro\watt} of cooling power at a base temperature of \SI{100}{\milli\kelvin}~\cite{bluefors_website}.
The fridge will allow operation well below the quantum noise temperature of \SI{720}{mK} at \SI{15}{\GHz}.

Although the fridge and magnet will be independently cooled by separate cold heads, they will be radiatively coupled through a shared cryogenic vacuum space at $\approx$\SI{4}{\kelvin}.
To protect the resonator from the $4\mathrm{K}$ thermal radiation of the magnet, the still shield ($\approx$\SI{1}{\kelvin}) of the dilution fridge will be extended into the main bore of the magnet.
This maximizes the usable volume inside of the magnet relative to a room temperature bore by reducing the number of redundant thermal shields needed around the resonator.
The shield extension will be made of \SIadj{1}{mm}-thick OFHC copper-plated SS (to minimize eddy current-induced damage in the event of a magnet quench) with an outer diameter of \SI{196}{mm}.
The shield extension will be fixed in the center of the magnet’s bore via a centering pin to maintain a \SI{3}{mm} gap between the still shield and the inner diameter of the magnet’s bore. 

\subsection{Resonator}
\label{sec:cavity}

\begin{figure*}[t!]
    \centering
    \includegraphics[width=\linewidth]{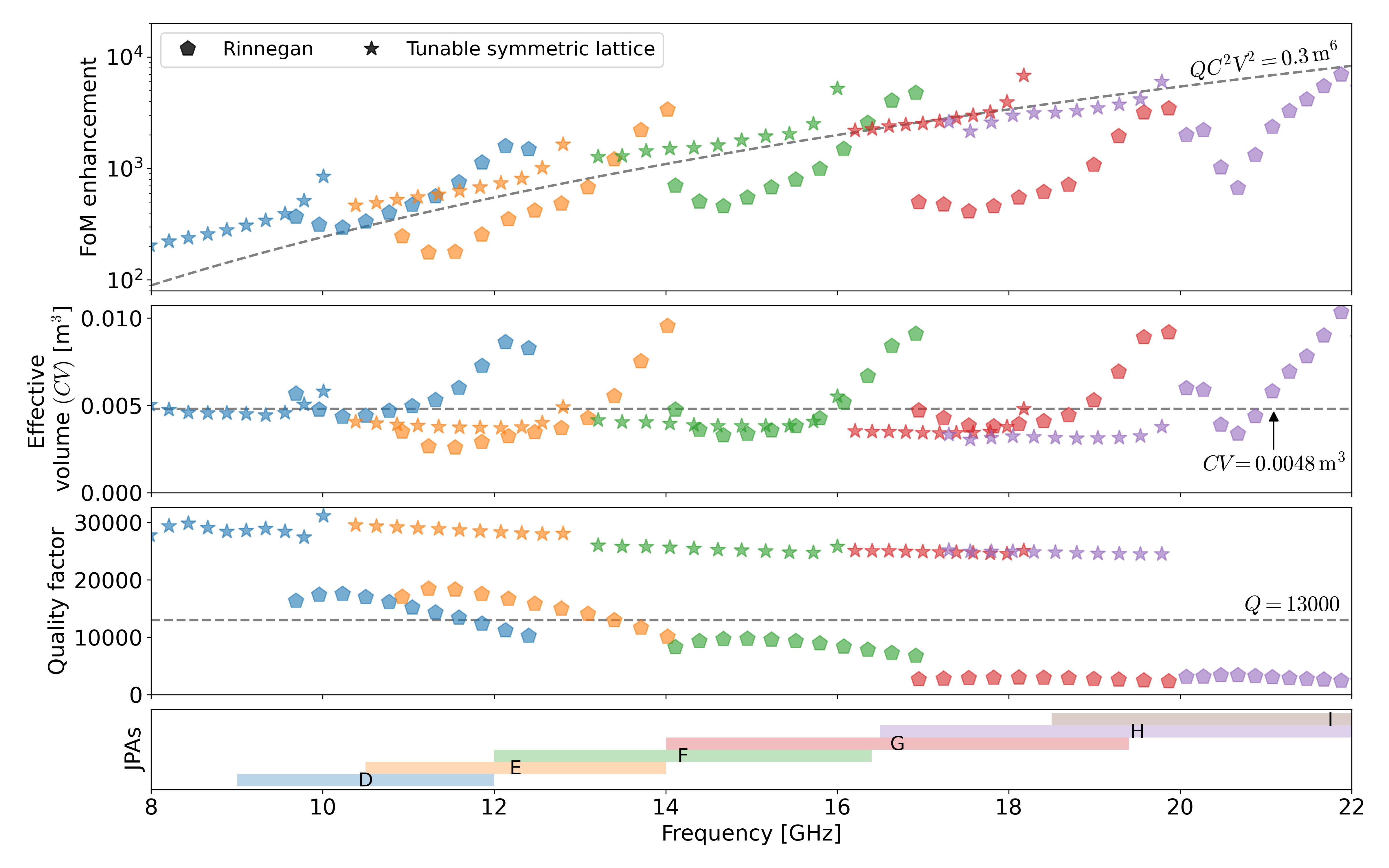}
    \caption{Performance of tunable wire-metamaterial resonators for the ALPHA Phase I experiment.
    (Top three panels) Simulated FoM enhancement, effective volume ($CV$), and cryogenic quality factor versus frequency for two tuning mechanisms: the Rinnegan (pentagons) and the tunable symmetric lattice (stars).
    Enhancement factors are calculated relative to a reference single cylindrical resonator, assuming a fixed resonator height of \SI{500}{mm} for all designs.
    Varying marker colors denote distinct structural configurations across the target spectrum.
    For comparison, the gray dashed lines represent the benchmark parameters ($QC^{2}V^{2} = \SI{0.3}{m^6}$, $CV = \SI{0.0048}{m^3}$, and $Q = 13000$), corresponding to the reference value in Equation~\ref{eq:scanrate}.
    (Bottom panel) Design frequency coverage bands for the corresponding suite of JPAs, which dictate the operational read-out ranges for the searches.}
    \label{fig:tuning_examples}
\end{figure*}

The ALPHA Phase I resonators, designed for the 10--\SI{20}{\GHz} range, represent an extrapolation of current resonator designs in dark matter axion experiments.
The design of the plasma haloscope uses a metamaterial with a plasma frequency determined not only by the average electron density of the array but also by a dramatically increased effective electron mass due to the mutual inductance of the wire array~\cite{Pendry_1998,Belov:PhysRevB:2003,Lawson:2019brd}.
To achieve this, the Phase I structures utilize movable metal tuning rods with spacing on the order of a centimeter and can be modeled using conventional microwave simulation codes, where the effective plasma frequency corresponds to the frequency of the lowest transverse magnetic (TM) mode.
The following discussion focuses on the resonator designs currently under evaluation for Phase I of the ALPHA experiment.

The development of practical tunable structures prioritizes sensitivity to dark matter axions, broad frequency coverage, and mechanical simplicity.
A viable tuning mechanism must dynamically alter the effective plasma frequency while preserving the key metrics that drive the experiment's scanning rate.
To maximize the scanning rate governed by Equation~\ref{eq:scanrate}, the figure of merit for these tunable designs is defined as $\mathrm{FoM} = Q C^{2} V^{2}$, which consists solely of resonator-dependent parameters.
To evaluate the efficiency of a given resonator design, we compare its FoM with that of a cylindrical cavity tuned to the same resonant frequency by adjusting its radius, defining the FoM enhancement as the ratio between the two.

\begin{figure}[t!]
    \centering
    \includegraphics[width=\linewidth]{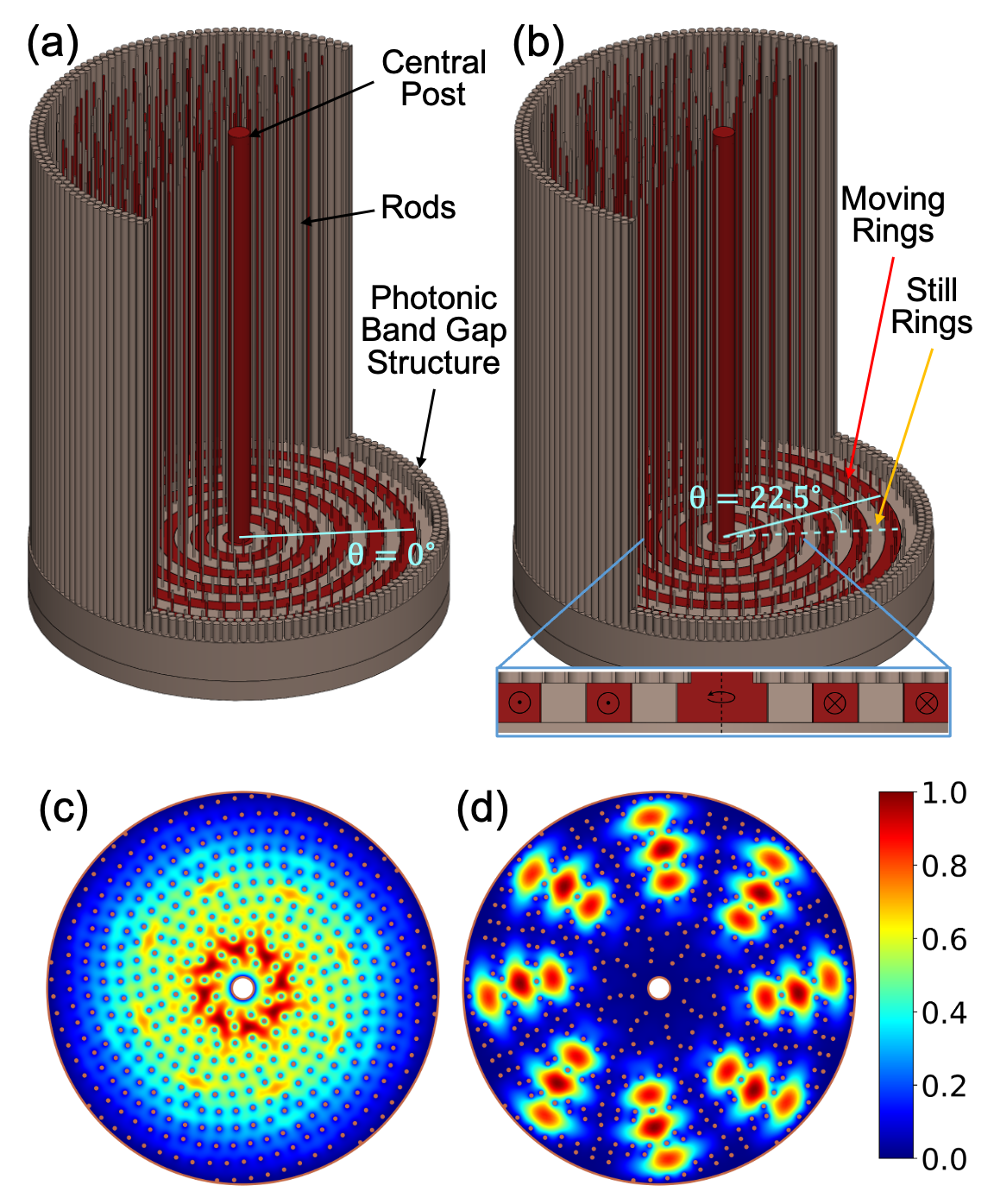}
    \caption{\label{fig:rinnegan_example}Example of a Rinnegan design that can cover the 10--\SI{12}{\GHz} range, utilizing a photonic band gap enclosure.
    The \SI{1}{mm}-diameter rods are arranged in a spiral configuration to efficiently fill the circular cross-section.
    Resonant frequency tuning is achieved by rotating the rods on the moving rings relative to the still rings.
    The thick central post ensures the simultaneous rotation of the top and bottom rings while maintaining the figure of merit of the design.
    3D cutaway models illustrate the assembly at rotation angles of (a) $\theta = 0^{\circ}$ and (b) $\theta = 22.5^{\circ}$.
    The electric field distributions for the lowest TM mode are shown for rotation angles of (c) $\theta = 0^{\circ}$ ($\nu=$~\SI{12.4}{\GHz}) and (d) $\theta = 22.5^{\circ}$ ($\nu=$~\SI{9.69}{\GHz}), where red and blue indicate high and low field intensities, respectively.}
\end{figure}

\begin{figure}[t!]
    \centering
    \includegraphics[width=\linewidth]{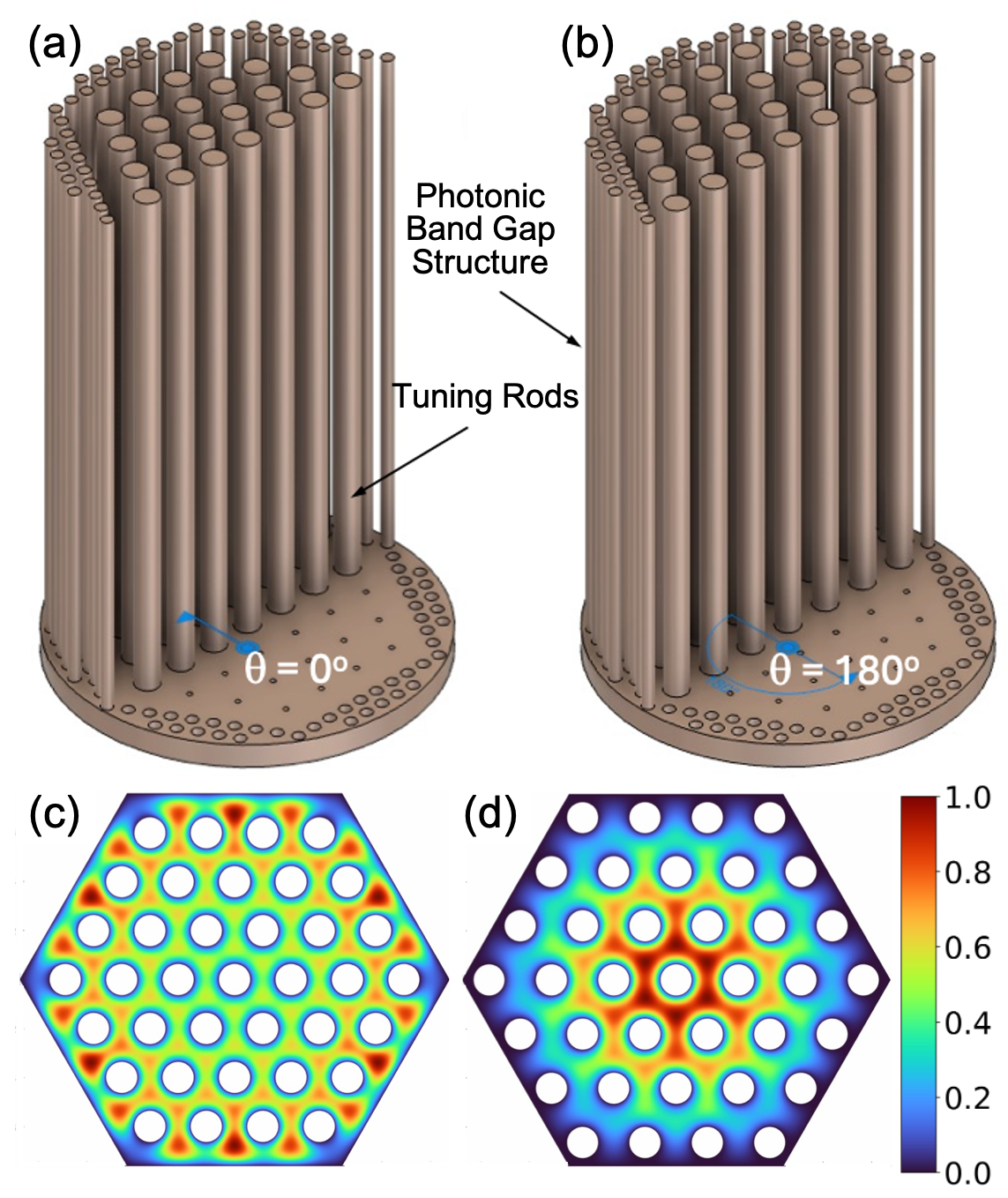}
    \caption{Example of a tunable symmetric lattice, with a photonic band gap enclosure for suppression of transverse electric (TE) modes. The outer diameter is 191 mm. Tuning is accomplished by pivoting thirty-seven metal rods of 12.7 mm diameter in unison. By displacing the axle of each tuning rod by an offset proportional to its radius from the axis of the resonator, triangular symmetry is maintained for all angles of rotation. (a) The rotation angle $\theta = 0^\circ$, corresponding to the minimum lattice constant, $a = $~\SI{23.0}{\mm}, and frequency $\nu =$~\SI{11.36}{\GHz} for this configuration. (b) The rotation angle $\theta = 180^\circ$, corresponding to the maximum lattice constant, $a =~$\SI{25.5}{\mm}, and frequency $\nu =~$\SI{9.17}{\GHz} for this configuration. The electric field associated with the TM$_{010}$ mode at the rotational angle of (c) $0^\circ$, and (d) $180^\circ$.}
    \label{fig:Berkeley:A}
\end{figure}

ALPHA aims to use multiple JPAs to continuously scan over a frequency range from \SIrange{10}{20}{\GHz}.
To do so, we aim to deploy several resonators, each tackling a several GHz subset of the full range that broadly aligns with the JPA coverage (see Figure~\ref{fig:tuning_examples}).
Because the system allows for the easy replacement of a resonator within the magnet, each resonator can be strategically optimized for its frequency coverage, operational ease, and figure of merit.

Metamaterial-based resonators face two central challenges: 1) tuning many rods simultaneously without resorting to overly complex mechanisms, and 2) maintaining a high quality factor despite the large number of rods employed in these geometries.
While there are many ways to tune plasma haloscopes~\cite{Lawson:2019brd, Balafendiev:arXiv:2025, CAPP:PRD:2023, Simanovskaia:RSI:2021, Goulart:RSI:2025, Lindahl:PRAppl:2026}, this phase of ALPHA focuses on two complementary designs: the ring nested group rotation (Rinnegan\footnote{RINg NEsted Group rotAtioN}) and the tunable symmetric lattice. As metamaterial-based resonators are a new paradigm for the axion haloscope, two independent designs are being pursued to further ensure success. These two designs address this trade-off directly; one prioritizes mechanical simplicity and ease of tuning, while the other adopts a more intricate multi-axis mechanism capable of achieving higher quality factors.
These designs are complementary in maximizing the figure of merit across different frequency bands, providing the flexibility to optimize the overall Phase I scanning strategy.
Furthermore, both designs are capable of tuning within the target frequency range and conserving longitudinal symmetry within the active volume of the resonator to minimize mode mixing.

The frequency coverage and figure of merit for these two tuning designs are illustrated alongside the JPA coverages in Figure~\ref{fig:tuning_examples}.
The resonator's inner radius is constrained to \SI{89}{mm} to fit within the ALPHA magnet after accounting for the wall thickness.
Although the magnet bore can accommodate a total height of \SI{1000}{mm}, the resonator inner height is set to \SI{500}{mm} for the initial Phase I configuration, allowing for a future upgrade to a two-stack multiple-resonator system.
Compared to a single cylindrical resonator scaled to the target frequency by reducing its radius while maintaining its height, the figure of merit of our proposed designs is two to three orders of magnitude higher.

The Rinnegan design adapts the spiral packing strategy discussed in \cite{Lindahl:PRAppl:2026} to achieve a high filling factor within a solenoid magnet while preserving longitudinal symmetry.
To avoid introducing intrusive mounting structures into the active volume, the spiral-packed rods are grouped into flush, concentric rings.
Tuning is achieved by rotating the even-numbered rings relative to the odd-numbered ones, which transitions the internal geometry from a spiral to a multiple-cell configuration~\cite{Jeong:PLB:2018}, effectively lowering the resonant frequency (see Figure~\ref{fig:rinnegan_example}). 
A major advantage of this mechanism is that it allows the entire wire-metamaterial to be tuned about a single rotational axis. Increasing the target frequency requires a smaller rod radius; at a fixed radius, the field localizes near the edge of the spiral as the frequency increases, degrading the form factor. Thinner rods suppress the localization and keep the effective volume high, but raise the surface ohmic loss, lowering the baseline quality factor. The Rinnegan design consequently trades quality factor for effective volume as the frequency increases, which is visible in Figure~\ref{fig:tuning_examples} as a declining quality factor at higher frequencies.

The other design for the initial ALPHA Phase I resonator is based on a triangular lattice of metal tuning rods, with tuning accomplished by the rods pivoting in unison around off-center axles.
Displacing the axle from the center of each rod by an offset proportional to its radial distance from the axis of the resonator ensures perfect triangular symmetry is maintained for all rotation angles; only the lattice constant and thus frequency changes (see Figure~\ref{fig:Berkeley:A}).
This design builds on the experience of the resonator currently deployed in HAYSTAC~\cite{Simanovskaia:RSI:2021,Goulart:RSI:2025} by simply extending the number of tuning rods from 7 to 37 for the lowest frequency resonator in the Phase I series.

To minimize the effects of mode mixing, both tuning methods incorporate a photonic bandgap (PBG) structure.
As the search frequency rises, the risk of interference from high-frequency transverse electric (TE) modes increases.
This interference mixes the electric fields of the desired mode with the TE modes, thereby degrading the form factor.
To mitigate this issue, PBG structures are integrated into the resonator perimeter to suppress TE mode confinement~\cite{Goulart:RSI:2025}.

\subsection{Quantum Readout}
\label{sec:quantumro}

Signals from the resonator will be read out via the antenna port and routed to a JPA~\cite{lehnert_jpa}, composed of an array of DC-SQUIDS, which will act as the first stage cryogenic amplifier. The principal challenge to operating a JPA above \SI{10}{\GHz} is the relatively low plasma frequency $\omega_{\rm J} = 1/\sqrt{L_{\rm J} C_{\rm J}}$ of the Josephson junctions that form the DC-SQUIDS, where $L_{\rm J}$ and $C_{\rm J}$ are the junction inductance and capacitance, respectively. In the Nb/Al-Ox/Nb trilayer process used in the HAYSTAC experiment~\cite{haystac_pI_design,brubaker_thesis}, $\omega_{\rm J} \approx{2\pi\times}$\SI{40}{\GHz}. The plasma frequency can be increased by thinning the oxide barrier, thereby increasing the critical current and in turn decreasing the inductance of each junction. The total inductance of the array can then be compensated simply by arraying more DC-SQUIDS.

Each device is designed with a target resonance frequency at zero applied magnetic flux. Applying a DC magnetic flux through the SQUID shifts this resonance frequency. We provide the flux by mounting the JPA inside a superconducting current loop, allowing the resonance to be tuned over a fractional frequency range of approximately 45\% of its center frequency.  To guarantee full coverage of the \SIadj{10}{\GHz} window that ALPHA Phase I is targeting we have designed JPAs with operating frequencies between \SI{9}{\GHz} and \SI{25}{\GHz} (see Figure~\ref{fig:tuning_examples}). The bands shown in Figure~\ref{fig:tuning_examples} are design targets. As discussed below, the first fabricated devices measured $\sim$20\% below design; the remaining devices are being fabricated with corrected junction parameters to recover the designed coverage. The designed bands take into account both the limited tuning range of each device and the variability in the design and fabrication, which results in some uncertainty in predicting the operating frequencies for a JPA design.

\begin{figure}[t!]
    \centering
    \includegraphics[width=\linewidth]{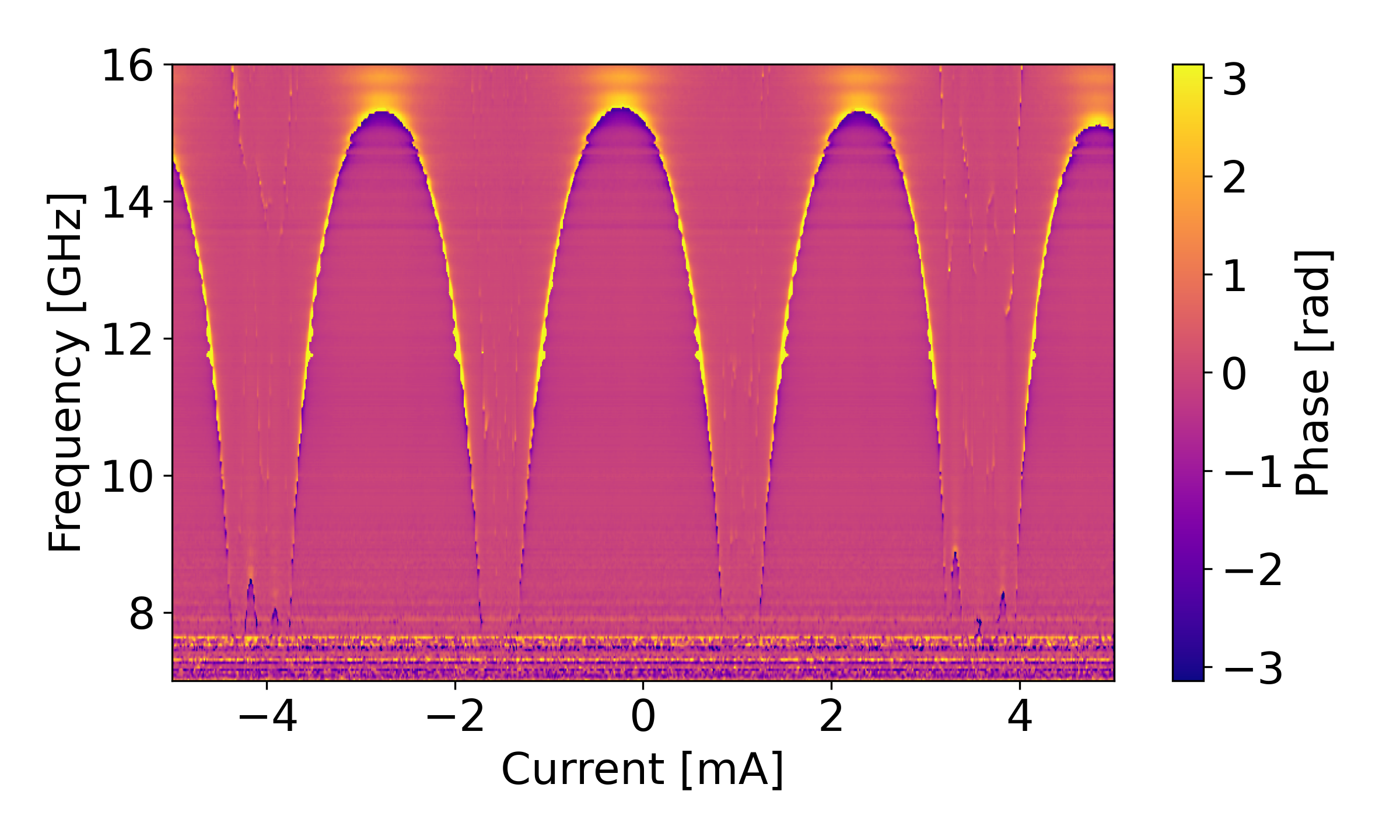}
    \caption{JPA~G frequency tuning curve over multiple periodic flux quanta. Current through the superconducting flux bias coils surrounding the JPA is swept from \SIrange{-5}{5}{mA}. At each flux bias point a VNA takes a measurement of the reflection off of the JPA from \SIrange{7}{16}{\GHz}. The JPA resonance frequency can be identified by the phase step of $\pi$ over the $\simeq100$~MHz BW of the JPA. The frequency response of the JPA is periodic with a period of 1 flux quantum, $\Phi_0$, corresponding to $\simeq2$~mA set by the mutual inductance between the SQUIDs on the JPA and the flux bias coils. At higher periods the array of SQUIDs become slightly dephased producing the artifacts visible around \SI{3.75}{\mA}. Normal operation is on the first branch from \qtyrange{-1}{1}{\mA}. The measured operation frequency is lower than the designed frequency by $\sim$20\%, which we attribute to a lower Josephson junction critical current $I_c$ and a higher total capacitance in simulations than expected. The remaining devices are being fabricated to account for all device capacitance sources and this reduced $I_c$, and will yield tuning ranges closer to the design.
    }
    \label{fig:JPA_G}
\end{figure}

Parametric gain is achieved by modulating, via an on-chip flux bias, the nonlinear inductance of the DC-SQUID array at twice the array's resonance frequency. This allows for amplification of weak signals within the JPA's bandwidth, with a tradeoff between gain and bandwidth set by the gain-bandwidth product $GB \equiv \sqrt{G}\,\Delta\nu$, which is approximately equal to the linewidth of the JPA's bare resonance. Throughout, $G$ denotes the two-quadrature (phase-insensitive) power gain referred to the signal sideband; this is the quantity returned by a vector-network-analyzer measurement of the pumped JPA, and the quantity that enters the gain-bandwidth product.  For the single-quadrature readout adopted here (Section~\ref{sec:daq}) the corresponding gain in the amplified quadrature is $G_{1Q} = 2G - 1 + 2\sqrt{G(G-1)} \simeq 4G$, i.e.\ $\sim 29$~dB for the $23$~dB operating point quoted below.  It is $G_{1Q}$, and not $G$, that refers downstream noise to the input; see Appendix~\ref{app:noise}. Each JPA is designed with a resonator quality factor of $Q = 50$ at zero flux, giving resonator linewidths of 200--\SI{400}{\MHz} across the 10--\SI{20}{\GHz} operating range. A device operating at, for example, \SI{20}{\GHz} would thus have $GB \approx \SI{400}{\MHz}$ and provide \SI{20}{\dB} of signal gain ($10\times$ amplitude gain) over a bandwidth of \SI{40}{\MHz}. We plan to operate the JPAs at $\sim$\SI{23}{\dB} of signal gain over approximately \SI{28}{\MHz}. Typical frequency response demonstrated by JPA G is shown in Figure~\ref{fig:JPA_G}.

Because the JPAs are extremely sensitive to magnetic fields, they require operation in a region with near zero magnetic field.
This is balanced by the need to operate the JPAs as close to the resonator output as possible to limit microwave losses.
To produce this low-field region a multi-stage shielding scheme, similar to that used in HAYSTAC~\cite{haystac_phase1}, will be employed.
The first stage comes from an active field cancellation coil, in series with the magnet's main coil, designed to create a field which opposes the fringe field of the main coil at a distance of \SI{90}{cm} above the center of the main coil.
This cancellation produces a low field region spanning a cylindrical volume of diameter \SI{41}{mm} and length \SI{100}{mm} that is expected to maintain a residual field $<$\SI{50}{G} throughout.
To further reduce the field the JPA will be located in the center of three stacked persistent superconducting coils each composed of 100 turns of Nb wire following the HAYSTAC design~\cite{haystac_pI_design}.
Assuming there is no flux present when these coils become superconducting at \SI{9}{\K}, they will act to maintain zero flux at their center as the magnet is ramped up to its operating field.
The final shielding stage will be a multi-layer shielding can that from inside out is composed of niobium, Amumetal 4K, aluminum and Amumetal 4K.
These layers help modify the boundary conditions such that the field strength and gradients inside of the can are significantly reduced.

In addition to the JPA, the overall design of the microwave receiver chain for ALPHA Phase I borrows heavily from the HAYSTAC receiver~\cite{haystac_phase1,haystac_phaseIIa}. The layout closely follows HAYSTAC Phase~II with the squeezer JPA removed: the JPA is flux-pumped at twice the resonator center frequency, and the same tone, divided by two, serves as the mixer local oscillator. Signal and image sidebands at $\omega_p/2 \pm \delta$, where $\omega_p$ is the JPA pump frequency and $\delta$ is a small frequency offset, therefore both lie within the resonator bandwidth and are amplified as a correlated pair, so that the measurement is single-quadrature (1Q) homodyne, with the amplified quadrature recovered by rotating the two digitized channels in analysis (Section~\ref{sec:daq}). The main difference between the ALPHA Phase I receiver and the one employed in HAYSTAC is that the microwave components and pump lines must utilize higher tolerance microwave coaxial transmission lines and connectors---\SI{3.5}{\milli\metre} and high-precision SMA connectors for signal lines, and \SI{2.92}{\milli\metre} connectors for pump lines which require frequencies twice the signal frequency and must extend up to \SI{40}{\GHz}. The full electronics and data acquisition chain is shown in Figure~\ref{fig:rf_chain_daq} in Appendix~\ref{app:noise}.

Assuming both the resonator and receiver are operated at a physical temperature of \SI{100}{\milli\kelvin}, the total system noise ($N_{\rm sys}$) with a single JPA operating at the quantum limit is expected to be 1.32~quanta, where one quantum denotes a noise spectral density of $h\nu$ per unit bandwidth.  
The additional 0.32~quanta above the  standard quantum limit (SQL) of 1 quantum ~\cite{Lamoreaux:2013koa,Haus:1962,Caves:1982zz} largely comes from the microwave losses in the receiver between the output of the resonator and the JPA (\SI{0.95}{\dB}) which are estimated to contribute 0.25~quanta, with a further 0.07 quanta coming from the follow-on receiver chain, including components such as high-electron-mobility transistors (HEMTs), cabling, and circulators. A squeezed state receiver, as demonstrated in HAYSTAC Phase II~\cite{haystac_phaseIIa}, was considered. However, at 10--\SI{20}{\GHz} the higher losses in commercially available microwave components---particularly cryogenic circulators---make the achievable squeezing too small to warrant the added complexity.

\subsection{Data Acquisition}\label{sec:daq}
The search for the axion signal is characterized by a detection of excess power above noise over a narrow frequency range corresponding to the width of the axion lineshape \cite{Turner:1990qx}. 
For ALPHA, signals in the \SIrange{10}{20}{\giga\hertz} frequency range will be downconverted with a homodyne receiver, whose local oscillator is set to the resonator center frequency, to an intermediate frequency (IF) range of 0--\SI{10}{\mega\hertz} prior to digitization as described below and detailed schematically in Figure~\ref{fig:rf_chain_daq}.  A signal entering the resonator through axion-photon conversion or through insertion via the weak port is passed through a circulator and parametrically amplified as described in Section \ref{sec:quantumro}.  A HEMT amplifier at \SI{4}{\kelvin} will provide a second stage of amplification before passing through a third stage of amplification and downconversion to IF at room temperature.  The resulting voltage fluctuations from the receiver will be sampled with two channels of an Alazar ATS9462 16-bit PCIe digitizer \cite{ats9462spec} at a rate $\geq$\SI{20}{MS/s} to ensure that the Nyquist frequency is above the sensitive bandwidth of the readout chain ($2\Delta\nu_{cav} = 2\nu_{cav}/Q_{L}$ where $Q_L$ is the loaded $Q$ of the resonator). Samples will be recorded in segments of 10 ms per channel, giving a frequency resolution of \SI{100}{\Hz}. This resolution is fine enough to resolve the Maxwellian shape of the axion signal, which is expected to have a linewidth of $\sim$\SI{15}{\kHz} at \SI{15}{\GHz} \cite{Turner:1990qx}. Each segment from the two channels will be rotated to produce a 1-quadrature readout aligned with the JPA's amplified quadrature to optimize the signal-to-noise ratio.  This use of a two-channel measurement is easily adapted for future upgrades to a sub-quantum readout method such as a squeezed-state receiver~\cite{haystac_phaseIIa}.  The power spectral density~(PSD) of each segment will be computed and averaged for each tuning configuration. This process will be repeated for the desired observation time, and the final output will be the full, averaged PSD over the observation time.  Computation of the PSD of each segment has been parallelized with ongoing data collection via python multiprocessing libraries \cite{python_multiproc, python_concurrent} to achieve an effectively zero processing deadtime. 
An alternative data collection mode will also provide the option to strategically save full time-traces for time-varying high-resolution analyses \cite{admxhighres,axionsubstructure}.  This is not the default mode, since the \SI{80}{MB/s} data rate produced by two channels at \SI{20}{MS/s} is too much to save. 

Calibration of the noise performance as discussed in Section~\ref{sec:quantumro} and Appendix \ref{app:noise} is achieved through Y-factor measurements with loads thermalized to the mixing chamber and a variable temperature source (VTS).  Further calibration of the signal is achieved through selective use of a probe tone and the injection of a fake axion signal via the calibration lines~\cite{haystacfakeaxion}.
 
The pre-processed spectra will be stored on a local server with backup to remote cloud servers accessible by participating institutions.  
In order to maintain optimal performance as the experiment scans through different frequencies, electronics will be controlled with a centralized control system based on the Simons Observatory Control System (OCS)~\cite{Koopman:2020gkh,ocs_github}.  This will allow for interfacing with each device in the detector, allowing for control as well as detailed record keeping to be automated systematically across the large array of devices.  

\section{Projection}
\label{sec:projection}

\begin{figure}[t!]
    \centering
    \includegraphics[width =\linewidth]{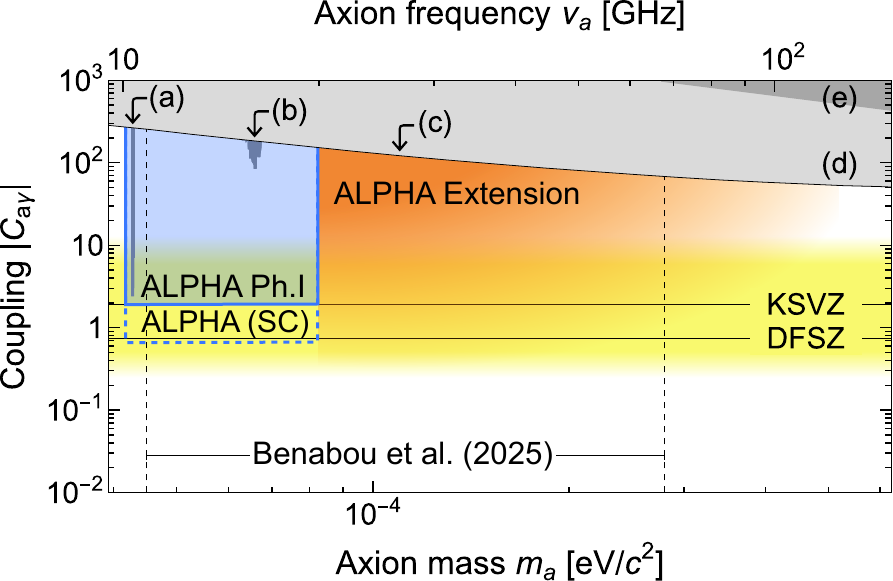}
    \caption{Discovery potential (see Section \ref{sec:projection}) for ALPHA Phase I (blue, solid) and in the case of a superconducting resonator (blue, dashed). The shaded orange area highlights the future possible range of the ALPHA extension to even higher frequencies, as discussed in Section~\ref{sec:discussion}.
    The plot also shows the current limits set by (a) QUAX~\cite{Alesini:2019ajt,Alesini:2020vny,QUAX:PRD:2022,QUAX:2023gop,QUAX:2024fut}; (b, c) ORGAN~\cite{McAllister:2017lkb,Quiskamp:2022pks,Quiskamp:2023ehr,Quiskamp:2024oet}; (d) NuSTAR~\cite{NuSTAR:PRL:2025}; (e) CAST~\cite{CAST:PRL:2024}, as well as a recent mass prediction~\cite{Benabou:2024msj}.}
    \label{fig:sensitivity}
\end{figure}

Figure~\ref{fig:sensitivity} shows the sensitivity projections for ALPHA Phase I, whose properties are summarized in Table \ref{tab:par_summary}. The current phase assumes 3.7 years of total integrated time covering the range 10--\SI{20}{\giga\hertz}. The reach of a future upgrade is also shown, both for Phase I itself, assuming a superconducting resonator (SC), and for future extensions of the experiment (ALPHA Extension).
In the latter case, the parameters describing the apparatus do not have a  clear path to implementation and the result should be regarded as a benchmark.

The computation follows the theoretical formalism and Monte Carlo approach of Ref.~\cite{GalloRosso:2022mhx}. 
The discovery potential shown in Figure~\ref{fig:sensitivity} is defined as the  coupling constant that would yield a median 5$\sigma$ local significant fluctuation.
It is worth noting that assuming the HAYSTAC~\cite{haystac:prd:2017,brubaker_thesis} protocol this discovery potential coincides with the expected 95\% confidence level (CL) upper limit for an ensemble of single scan experiments---as the condition triggering the remeasuring of possible signals (``rescanning'') is defined with respect to the discovery potential.
\footnote{A conventional frequentist upper limit would be defined from the observed data instead. It is the value of the coupling constant $|C_{a\gamma}|^{\text{\textsc{ul}}}$ for which the probability of obtaining data as compatible or less compatible with the signal hypothesis ($|C_{a\gamma}| > 0$) than the observed data is equal to $1-\mathrm{CL}$, under the assumption of that specific signal strength ($|C_{a\gamma}| = |C_{a\gamma}|^{\text{\textsc{ul}}}$).}
Further, the procedure to claim a discovery is not purely statistical, as a series of additional sanity checks on any candidate would have to be performed. These include the line shape of the signal, checks for RF sources outside of the experiment, dependence on the applied magnetic field (in the case of an axion) and possible daily modulation (in the case of a dark photon~\cite{Gelmini:PRD:2020,Caputo:2021eaa}).

For simplicity, we did not consider a rescanning protocol. Rescanning would on the one hand enhance the statistical significance of discovery at given frequency, on the other hand add to the integration time. Assuming a 3.36$\sigma$ threshold (corresponding to HAYSTAC protocol~\cite{haystac_phaseIIfull}) and tested frequencies spaced $\Delta\nu_a/3$ over the scanning range~\cite{brubaker_thesis} the expected number of noise fluctuations that would require a rescan is $\sim800$ over the 10--\SI{20}{\giga\hertz} range, and this would increase the integration time by only about 4\%.

\begin{table}[t]
\begin{tabular}{ll|c|c}
        \hline \hline 
        \multirow{2}{*}{Parameter}  & & \multicolumn{2}{c}{ALPHA}\\
        & & Normal & SC\\
        \hline \hline
        Quality factor & $Q$
        & \num{1.3e4} & $10^{6}$\\
        Form factor & $C$ 
        & \multicolumn{2}{c}{ 0.4 } \\
        Volume & $V$
        & \multicolumn{2}{c}{ \SI{0.012}{m^{3}} } \\
        Magnetic field strength & $\left|\mathbf{B}_{\rm ext}\right|$
        & \multicolumn{2}{c}{\SI{9}{\tesla}} \\
        Total system noise & $N_{\rm sys}$
        & \multicolumn{2}{c}{ 1.32 } \\
        Integration time & $\tau$
        & \multicolumn{2}{c}{ 3.7 y } \\
        \hline
    \end{tabular}  
    \caption{Summary table of the parameters used in the sensitivity projections, for ALPHA Phase I (yielding $QC^{2}V^{2} =$~\SI{0.3}{m^6}) and in the case of a superconducting resonator (SC).}
    \label{tab:par_summary}
\end{table}

\section{Future Enabling Technology}\label{sec:discussion}

Possible future extensions of ALPHA also include larger‑bore and higher‑field magnet facilities, as well as superconducting or hybrid resonator technologies capable of maintaining high quality factors at elevated frequencies and magnetic fields. These developments are broadly relevant to both Phase~I upgrades and potential future operation. By contrast, operation above $\sim$20~GHz unavoidably requires a departure from conventional readout schemes, making the development of new detection strategies a central enabling requirement for higher‑frequency searches.

\subsection{Superconducting Resonator}
Resonators designed to search for the post-inflation axion will face an increasingly unfavorable volume-to-surface ratio and thus diminished $Q$ as more conducting tuning elements are introduced to drive the frequency upwards.
This motivates exploration of superconducting thin films to improve $Q$, as recent research demonstrates $Q>10^{6}$ even in strong magnetic fields~\cite{ahn2023thesis}.
For ALPHA Phase I, operating up to \SI{20}{\GHz}, a resonator that maintains high $Q$ in a high magnetic field, $B \sim$~\SI{10}{T} would be beneficial.
For future phases, searching at frequencies up to \SI{50}{\GHz} will require wire-array metamaterials involving densely packed thin wires, $\sim$\SI{100}{\micro m} spaced by $\sim$\SI{1}{\mm}.

The Type-II superconductor magnesium diboride (MgB$_2$) is under active investigation at UC Berkeley and appears to be an attractive solution.
It possesses a high critical temperature, $T_c =$~\SI{39}{K}, and a high upper critical field, $H_{c2} =$~\SI{14}{T}, which can be increased by doping.
Furthermore, doping with carbon can also introduce pinning sites which will help suppress vortex dissipation at RF frequencies, as the axion search requires operation in a high magnetic field.

\begin{figure}[t!]
    \centering
    \includegraphics[width = \linewidth]{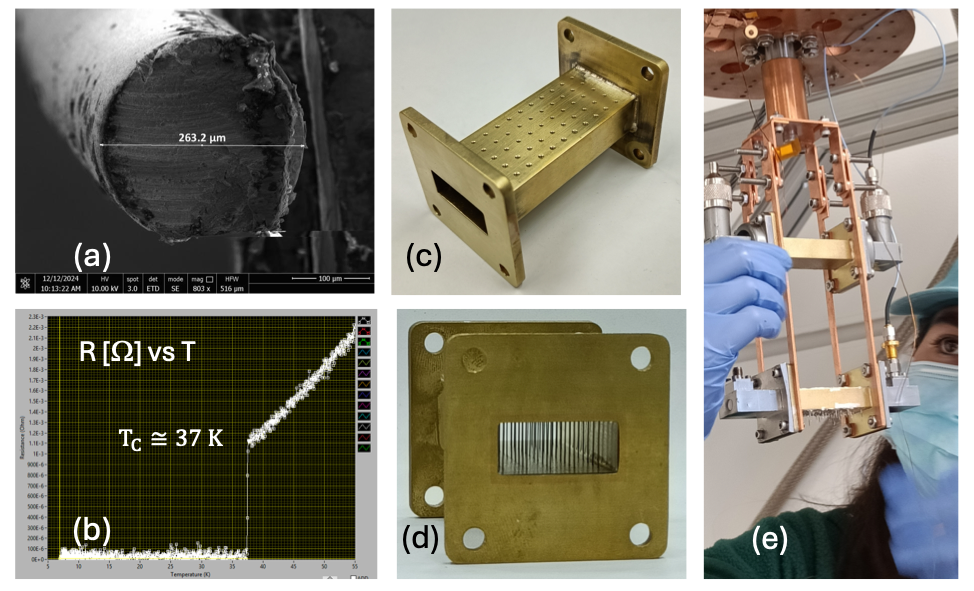}
    \caption{Waveguide-mounted superconducting wire-array metamaterials for tests.
    (a) Microphotograph of a niobium wire deposited with magnesium diboride of several microns thickness.
    (b) DC resistance R as a function of temperature $T$. After adjustments to the deposition process, $T_{c}$ was measured to be \SI{39}{\K}.
    (c) X-band waveguide made of brass with a lattice of 5 x 11 holes to mount the wires.
    (d). Waveguide loaded with a wire lattice.
    (e) Two waveguides mounted in the \SI{4}{\K} refrigerator; a copper wire array was always measured at the same time as with a superconducting wire array as a control.}
    \label{fig:sccavity}
\end{figure}

Initial studies have focused on the determination of superconducting wire-array metamaterial properties as a function of temperature $T$ and magnetic field $B$ by means of $S_{21}$ (transmission) measurements.
Following the early work of Ref.~\cite{Ricci:APL:2005}, a series of X-band waveguide-mounted wire arrays of Cu, Nb and MgB$_2$ were fabricated (Figure~\ref{fig:sccavity}).
These were square lattices of \SI{250}{\um} wires, with lattice constants of \SI{4.5}{\mm} and \SI{6.4}{\mm}.
The MgB$_2$ wires were made by a Hybrid Physical Chemical Vapor Deposition (HPCVD) process on niobium wires as substrates.
Preliminary tests at various $T$ with $B=0$ were analyzed with the same formalism as $S_{21}$ measurements of large open-frame wire arrays~\cite{Kowitt:PRAppl:2023,Wooten:AdP:2024}, and exhibited sharp drops in the loss at the expected $T_c$ for Nb (\SI{9}{\K}) and MgB$_2$ (\SI{39}{\K}).
These measurements will be repeated and extended up to $B = $~\SI{9}{\T}; a prototype MgB$_2$ photonic band gap resonator has also been made and its $Q$ will be mapped as a function of ($B$, $T$).

\subsection{ALPHA Future Readout}

The ALPHA Phase I experiment targets the 10--\SI{20}{\GHz} frequency range, where quantum‑limited Josephson parametric amplifiers provide a mature and effective readout solution. At significantly higher frequencies, however, Josephson‑junction‑based amplifiers become increasingly challenging to realize due to constraints on junction plasma frequency, loss, and device yield. As a result, extending axion searches beyond $\sim$20~GHz will require alternative measurement strategies.

One promising approach is frequency down‑conversion based on kinetic inductance nonlinearity, which remains operative at substantially higher frequencies. Future extensions of ALPHA are considering kinetic-inductance-based devices to mix high frequency axion signals back down to the low microwave range for which traditional superconducting quantum amplifiers are well suited. The device, currently being tested, can perform two-mode squeezing to add gain to the converted signal with a minimum of half a photon of added noise, or pure conversion with, in principle, no added noise. The down-converted signal can then be amplified using standard junction-based quantum measurement devices.

In parallel, fundamentally different readout paradigms are under active development for high‑frequency haloscopes, including single‑photon detection schemes such as Rydberg‑atom‑based detectors and superconducting qubit-based photon sensing~\cite{PhysRevD.109.032009,quax2025,qubitakash2021,junctionsinglephoton2022}. These approaches bypass the quantum measurement noise inherent to linear amplification and become increasingly attractive at higher frequencies where thermal backgrounds are suppressed~\cite{Lamoreaux:2013koa}.
ALPHA is designed to accommodate this range of future readout technologies, enabling a phased experimental program in which plasma haloscope resonators can be paired with different detection strategies as the experiment extends to higher axion masses.

\section{Conclusion}

In this paper we present the technical design for the ALPHA axion haloscope, with primary emphasis on the Phase~I experiment.
The experiment is designed to search for the post‑inflation axion with sensitivity approaching the KSVZ benchmark between 10 and \SI{20}{\GHz}.
Two resonator designs, the Rinnegan and the tunable symmetric lattice, are being developed in parallel; the final Phase I series will be assembled from whichever design gives the higher figure of merit in each sub-band.
These designs integrate multiple JPAs such that each resonator broadly aligns with the operational bandwidth of its respective JPA.
Phase~I paves the way for the next generation of ALPHA detectors, which would incorporate superconducting plasma resonators and alternative readout schemes.
These next generation detectors will extend the plasma haloscope approach to higher axion masses, enabling sensitivity deeper into the dark‑matter axion parameter space.

\begin{acknowledgments}
We gratefully acknowledge funding support from the Alfred P. Sloan Foundation, the Gordon and Betty Moore Foundation, the John Templeton Foundation, and the Simons Foundation. 
This material is also based upon work supported by the U.S.~Department of Energy, Office of Science, Office of High Energy Physics, QuantISED 2.0 program under Award Number DE-SC0025944.
This research was supported by the Swedish Research Council (VR) under Dnr.\ 2019-02337 ``Detecting Axion Dark Matter In The Sky And In The Lab'' (AxionDM). We gratefully acknowledge the funding support from the Knut and Alice Wallenberg Foundation and Olle Engkvists Foundation.
This research utilized the Sunrise HPC facility supported by the Technical Division at the Department of Physics, Stockholm University, \url{http://doi.org/10.5281/zenodo.TBD}.
E.B.~and C.L.~are supported by the National Science Foundation Graduate Research Fellowship under Grant No.~DGE-2139841.
J.E.G.\ gratefully acknowledges support from the University of Iceland Research Fund.
HAYSTAC is supported by the National Science Foundation under grants PHY-2514170, PHY-2514172, PHY-2209556, PHY-1306729, the Heising-Simons Foundation under grants 2014-0904, 2014-182, 2016-044, and the David and Lucile Packard Foundation grant number 2022-74683.
We gratefully acknowledge Prof.~Xiaoxing Xi and Dr.~Ke Chen of Temple University for providing  MgB$_2$ coated test pieces.
We gratefully acknowledge S.K. Lamoreaux for contributions in the early stages of the project.

\end{acknowledgments}

\appendix

\begin{figure*}[th!]
    \centering
    \includegraphics[width=0.9\textwidth]{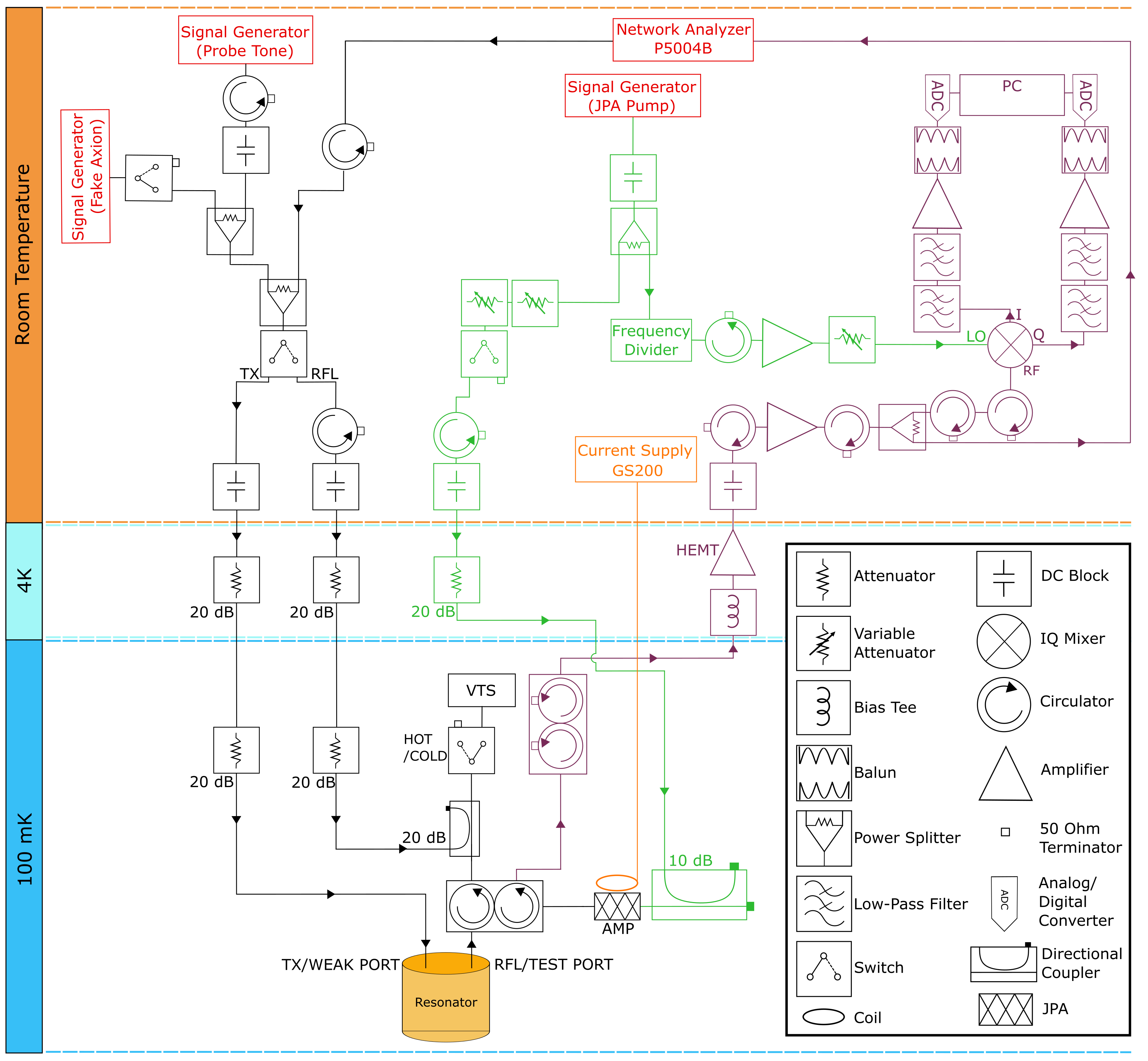}
    \caption{Data acquisition and electronics schematic diagram for the ALPHA experiment. (Black) The calibration equipment used to measure the resonator transmission (tx) and reflection (rfl) along with system noise. This calibration equipment can also inject CW signals for gain monitoring, along with ``fake axion" signals for analysis validation. (Green) High frequency 20--\SI{40}{\GHz} electronics used for JPA Pumping, and for the mixer local oscillator (LO) after passing through a divide by 2 circuit element. (Purple) Main receiver chain from the resonator out to the digitizer.}
    \label{fig:rf_chain_daq}
\end{figure*}

\section{Receiver Chain and Noise Calculation}
\label{app:noise}

Figure~\ref{fig:rf_chain_daq} follows the architecture established by the HAYSTAC receiver~\cite{haystac_pI_design, haystac_phase1, haystac_phaseIIa}, with three functional groups distinguished by color. The black calibration path allows signal injection on the resonator's two ports: the weakly coupled TX port, used for transmission (tx) measurements that locate the resonant frequency and determine the loaded quality factor, and the strongly coupled RFL/TEST port, used for reflection (rfl) measurements that determine the antenna coupling $\beta$. A room-temperature switch selects between the two so that both measurements share a single receiver chain. Each input line contains attenuators anchored at \SI{4}{\kelvin} and at the MXC, which thermalize the line and prevent room-temperature Johnson noise from reaching the resonator. The RFL calibration line enters through the \SI{20}{dB} coupled port of a directional coupler, with a VTS and a cold load attached to its through port. A double-junction circulator then routes this line into the RFL port of the resonator and directs the reflected signal onward to the JPA and the receiver.

The green path generates the \SIrange{20}{40}{\GHz} JPA pump. Because the JPA is flux-pumped at twice the resonator center frequency, dividing the pump by two produces a local oscillator (LO) phase-locked to the pump. This ensures that the JPA amplified quadrature is a fixed linear combination of the two digitized channels rather than one that drifts with the relative phase of two independent synthesizers. The pump and LO lines each have software-controllable variable attenuators installed at room temperature so their relative amplitudes can be individually optimized while maintaining phase lock. The pump is injected into the JPA through a \SI{10}{dB} directional coupler, and the DC flux setting the JPA resonance frequency is supplied by a low-noise programmable current source driving the superconducting flux-bias coil. On the resonator output, two double-junction circulators, each providing \SI{35}{dB} of isolation, prevent reflected noise from making its way back to the JPA and resonator. The purple receiver path carries the amplified signal to room temperature, where an IQ mixer downconverts it and the two quadratures are digitized on separate channels and rotated in analysis as described in Section~\ref{sec:daq}.

The total system noise ($N_{\rm sys}$) of the experiment is given by
\begin{equation}
    N_{\rm sys} = \frac{1}{e^{h\nu/k_{B}T_{\rm MXC}} - 1} + \frac{1}{2} + N_{A}
    \label{eq:Nsys}
\end{equation}
where the first two terms together give the noise spectral density of the resonator and the cold load~($N_{\rm MXC}$), held at a physical temperature $T_{\rm MXC}$ (a thermal occupation term plus the $1/2$ quantum of zero-point fluctuation), and the last term represents the input-referred added noise of the receiver itself.
With the LO centered at the half-pump frequency, the measured quadrature $\hat X = \tfrac{1}{2}[\hat a(\delta) + \hat a^\dagger(-\delta)]$ combines the output from both sidebands at $\omega_{\rm p}/2 \pm \delta$.
Its noise therefore contains the thermal and zero-point contributions from both sidebands, while an axion signal occupies only one.
This is the origin of the factor of $2$ in the noises of Equation~\ref{eq:Nsys_cascade}, as will be shown below.
Because $N_{\rm sys}$ is normalized to the single-sideband signal in this way, it should be compared directly to the full axion power $P_{\rm ax}$ in Equation~\ref{eq:scanrate} not to $P_{\rm ax}/2$, as in HAYSTAC's convention~\cite{haystac_phaseIIa,haystac_phaseIIb,haystac_phaseIIfull} where the quoted noise number is halved instead.

Additional contributions to $N_{A}$ arise from both added noise and losses from the other components necessary to transmit and amplify signals to the room-temperature readout. By treating transmission losses to the signal power as effective increases to the noise, the total noise can be determined by cascading each element, accounting for both its added noise and loss. Combining these noise sources into Equation \ref{eq:Nsys} yields Equation \ref{eq:Nsys_cascade}.
\begin{equation}
\begin{split}
    N_{\rm sys} &= \frac{1}{\lambda}2N_{\rm MXC} + \frac{1-\eta}{\eta\lambda G_{1Q}}2N_{\SI{4}{K}} + \frac{1}{\eta\lambda G_{1Q}} 2 N_{H}\\
    &\approx 1.32
    \label{eq:Nsys_cascade}
\end{split}
\end{equation}
where $G_{1Q}$ is the single-quadrature signal gain of the JPA, assumed to be \SI{29}{\dB}.
The transmission efficiency between the resonator and JPA ($\lambda$) is estimated to be 0.8, with a \SI{0.86}{\dB} loss coming from the circulators needed to prevent noise from leaking back into the resonator from the JPA.
The remaining loss comes from the $\sim$\SI{1}{\meter} of NbTi superconducting coax cabling needed to route signals from the resonator to the JPA, assumed here to be \SI{0.045}{\dB\per\meter}. Following that, we estimate a transmission efficiency ($\eta$) of 0.44 between the JPA and the HEMT, which sources noise between MXC temperature and \SI{4}{\kelvin} ($N_{\SI{4}{K}}$). Finally, from its specifications, the HEMT\footnote{Low Noise Factory \texttt{LNF-LNC6\_20D}} ($N_{H}$) is expected to add an additional 7~quanta (\SI{5}{\kelvin}) of noise at \SI{15}{\GHz}. From just these components the total noise is expected to be $N_{\rm sys} \approx 1.32$~quanta, with 1.25~quanta coming from the first term representing the noise and the losses prior to the JPA, and the remaining 0.07~quanta from the follow-on chain.

The largest uncertainty in this calculation comes from the superconducting cabling that runs through the inside of the magnet, whose loss is expected to be higher than that of SC cabling operated in zero field. There is not a great deal of literature for the expected loss at these frequencies in large fields, with the existing literature having contradictory predictions of frequency scalings between $\nu^{0.5}$ and $\nu^{2.7}$. For this calculation it is assumed that the loss scales as $\nu^{1/2}$, giving \SI{0.045}{\dB\per\meter} at \SI{15}{\GHz}. In the case that the scaling instead follows $\nu^{2.7}$, the loss would increase to \SI{1.83}{\dB\per\meter}. This would result in a 50\% increase in noise over the baseline assumption, giving 2.0~quanta of total noise.

A squeezed-state receiver, as demonstrated in HAYSTAC Phase~II~\cite{haystac_phaseIIa}, was considered for ALPHA Phase~I but ultimately not pursued. In a squeezed-state receiver configuration, a squeezer JPA (SQZ) generates a squeezed vacuum state that is routed to the resonator input, while a second amplifier JPA (AMP) reads out the output. The achievable squeezing $S$ is set primarily by the inter-JPA loss $\eta$:
\begin{equation}
    S = \frac{\eta}{G_{\mathrm{SQZ}}} + 1 - \eta
\end{equation}
where $G_{\mathrm{SQZ}}$ is the single-quadrature gain of the squeezer. The dominant loss between the JPAs comes from the triple-junction circulator, which at ALPHA's operating frequencies carries a quoted insertion loss of \SI{1.3}{\dB}—more than three times the \SI{0.4}{\dB} loss of the equivalent HAYSTAC component. Possible non-negligible NbTi coaxial cable losses in high magnetic field discussed above further compound this. Accounting for these components, the predicted inter-JPA loss lies between an ideal-case \SI{2.47}{\dB} and a realistic-case \SI{3.78}{\dB} (including an additional \SI{1.32}{\dB} of uncharacterized loss inferred from HAYSTAC experience), yielding squeezing bounds of $S_{\mathrm{ideal}} = -3.35$~dB and $S_{\mathrm{realistic}} = -2.2$~dB, compared to $\sim$\SI{-4}{\dB} achieved on HAYSTAC. These correspond to scan-rate enhancements of only 1.43$\times$ to 1.75$\times$. Furthermore, HAYSTAC experience shows that squeezing tends to vary over the full bandwidth of the experiment, so the ideal-case figure would likely not be achievable at all frequencies. Given the modest and uncertain gain in scan rate relative to the added complexity, cost, and maintenance burden of a squeezed-state receiver, squeezing was not incorporated into the ALPHA Phase~I design; however, the electronics were designed to allow for a two-JPA squeezed-state upgrade path.

\bibliography{main}

\end{document}